\documentclass[aps,prb,twocolumn,superscriptaddress,groupaddress,10pt]{revtex4-1}
\usepackage[dvipsnames]{xcolor}
\usepackage{ulem}

\usepackage{graphpap}
\usepackage{graphicx}
\usepackage{hyperref}
\usepackage{wasysym} 
\usepackage{comment}

\begin{document}
	
\title{Quantum transport through MoS$_2$ constrictions defined by photodoping}

\author{Alexander Epping}
\affiliation{These authors contributed equally to this work}
\affiliation{JARA-FIT and 2nd Institute of Physics, RWTH Aachen University, 52074 Aachen, Germany}
\affiliation{Peter Gr\"unberg Institute (PGI-9), Forschungszentrum J\"ulich, 52425 J\"ulich, Germany}

\author{Luca Banszerus}
\affiliation{These authors contributed equally to this work}
\affiliation{JARA-FIT and 2nd Institute of Physics, RWTH Aachen University, 52074 Aachen, Germany}
\affiliation{Peter Gr\"unberg Institute (PGI-9), Forschungszentrum J\"ulich, 52425 J\"ulich, Germany}

\author{Johannes G\"uttinger}
\affiliation{JARA-FIT and 2nd Institute of Physics, RWTH Aachen University, 52074 Aachen, Germany}
\affiliation{Peter Gr\"unberg Institute (PGI-9), Forschungszentrum J\"ulich, 52425 J\"ulich, Germany}

\author{Luisa Kr\"uckeberg}
\affiliation{JARA-FIT and 2nd Institute of Physics, RWTH Aachen University, 52074 Aachen, Germany}

\author{Kenji Watanabe}
\affiliation{National Institute for Materials Science, 1-1 Namiki, Tsukuba, 305-0044, Japan }

\author{Takashi Taniguchi}
\affiliation{National Institute for Materials Science, 1-1 Namiki, Tsukuba, 305-0044, Japan }

\author{Fabian Hassler}
\affiliation{JARA-Institute for Quantum Information, RWTH Aachen University, 52074 Aachen, Germany}

\author{Bernd Beschoten}
\affiliation{JARA-FIT and 2nd Institute of Physics, RWTH Aachen University, 52074 Aachen, Germany}

\author{Christoph~Stampfer}
\affiliation{JARA-FIT and 2nd Institute of Physics, RWTH Aachen University, 52074 Aachen, Germany}
\affiliation{Peter Gr\"unberg Institute (PGI-9), Forschungszentrum J\"ulich, 52425 J\"ulich, Germany}

\date{ \today}

\begin{abstract}
  We present a device scheme to explore mesoscopic transport through molybdenum disulfide (MoS$_2$) constrictions using photodoping. The devices are based on van-der-Waals heterostructures where few-layer MoS$_2$ flakes are partially encapsulated by hexagonal boron nitride (hBN) and covered by a few-layer graphene flake to fabricate electrical contacts. Since the as-fabricated devices are insulating at low temperatures, we use photo-induced remote doping in the hBN substrate to create free charge carriers in the MoS$_2$ layer. On top of the device, we place additional metal structures, which define the shape of the constriction and act as shadow masks during photodoping of the underlying MoS$_2$/hBN heterostructure. Low temperature two- and four-terminal transport measurements show evidence of quantum confinement effects.
\end{abstract}

\maketitle

%%%%%%%%%%%%%%%%%%%%%%%%%%%%%%%%%%%%%%%%%%%%%%%%%%%%%%%%%%%%%%%%%%%%%
%% Start the main part of the manuscript here.
%%%%%%%%%%%%%%%%%%%%%%%%%%%%%%%%%%%%%%%%%%%%%%%%%%%%%%%%%%%%%%%%%%%%%

\section{Introduction}
Van-der-Waals heterostructures based on graphene and transition metal dichalcogenides (TMDCs) are attracting increasing attention in mesoscopic physics and solid-state research\cite{Novo16,Geim13,Rad10,Herr13,Rus08,Ter16}. One of the main challenges in this field is to find ways of confining charge carriers in well defined device geometries. Achieving a high degree of control over carrier confinement is necessary for manipulating individual charges or spins\cite{Elz04,Wee88,Wep13,Joh92}, as well as for investigating mesoscopic physics phenomena such as quantized conductance or valley filtering in bilayer graphene and TMDCs \cite{Ryc07,Piso17}. Moreover, good control over carrier confinement is a necessary requirement for studying fundamental material properties such as the Land\'{e} g-factor and the charge carrier's effective mass in mesoscopic devices.    

Because of the absence of a band gap, mesoscopic physics in graphene is mostly studied in devices where the material has been etched into the desired shape\cite{Ter16,Rus08}. These devices typically suffer from scattering and trap states on their rough edges\cite{Ter16,Tom11,Bis14}. In contrast, bilayer graphene\cite{Gos12,All12} and semiconducting \mbox{TMDCs}\cite{Son15,Ke16} allow for soft electrostatic confinement with the help of metallic gates, similarly to what has been pioneered in two-dimensional electron gases (2DEGs) in conventional semiconductor heterostructures for the past decades\cite{Wee88,Gold1998,Petta05}. First experiments on confining charge carriers in MoS$_2$ and forming one-dimensional transport channels by electrostatic gating have recently been performed \cite{KLee16,Ke16,Piso17,ZZZhang17,CHSharma17,SBhandari17}. 
Here, we demonstrate an interesting method of defining arbitrary doping profiles in van-der-Waals heterostructures based on few-layer (two to five layers) MoS$_2$ encapsulated in hexagonal boron nitride (hBN). In analogy to remote doping in semiconductor 2DEGs, we make use of the photodoping effect recently observed in hBN/graphene heterostructures\cite{Ju14,Wong15,Neu16}. We use this effect to define stable and smooth sub-wavelength lateral doping profiles in MoS$_2$ employing metal shadow masks that are lithographically defined on top of an hBN/MoS$_2$/hBN heterostructure. Our photo-induced remote doping approach enables us to reach free charge carrier densities of up to $10^{13}$~cm$^{-2}$, when using an additional gate voltage during light illumination. It thus allows to create arbitrary carrier density profiles defined by the shadow mask geometry. The feasibility of our approach is demonstrated by studying quantum transport through constrictions in MoS$_2$. We expect that our method allows to fabricate more complex geometries such as antidot lattices or Aharonov-Bohm rings with the advantage that the metal used for shadow masking does not need to be electrically contacted, when studying mesoscopic transport.

\begin{figure*}[t!]\centering
	\includegraphics[draft=false,keepaspectratio=true,clip,%
	width=\linewidth]%
	{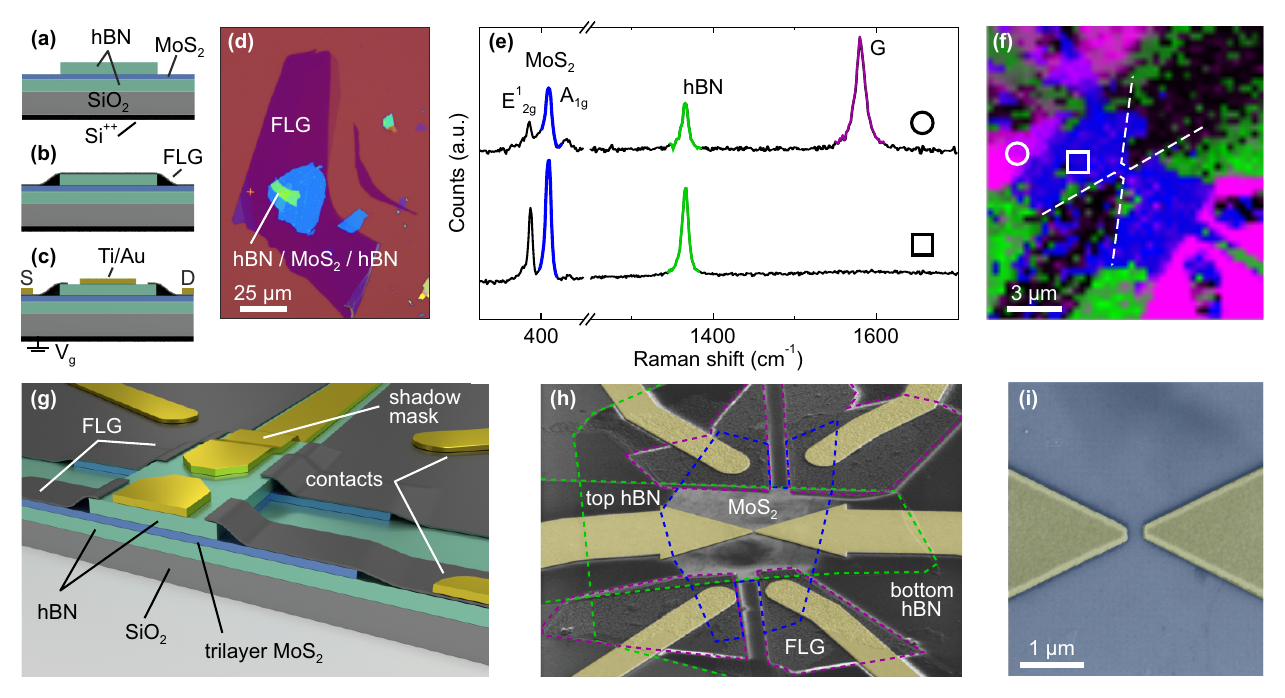}
	\caption[Fig01]{ (a)-(c) Schematic illustration of the device fabrication. (a) MoS$_2$ is partially encapsulated by two hBN flakes. (b) A few-layer graphene (FLG) flake is transferred on top of the entire heterostructure. (c) After etching the FLG into the desired shape, the shadow mask structure and the source (S) and drain (D) contacts are made. The Si$^{++}$ substrate is used to apply the gate voltage V$_g$. (d) Optical microscope image of the complete van-der-Waals heterostructure. (e) Raman spectra on a final device recorded in the lead area indicated by the white circle in panel f (top) and in the sandwiched part of the device indicated by the white square in panel f (bottom) showing the individual Raman peaks of hBN (green), graphene (purple) and MoS$_2$ (blue). (f) Raman map showing the integrated intensities of the hBN peak, the G peak of the FLG and the A$_{1g}$ mode of the MoS$_2$, represented by the colors purple, blue and green, respectively. The black areas are covered by metal whose boundaries are indicated by the white dashed lines. (g) Schematic of a typical multi-terminal device (Si$^{++}$ back gate not shown). (h) Scanning electron microscopy image of a four-terminal device. The individual flakes are highlighted by dashed lines with the same color as in panel e. (i) Close-up of the shadow mask that forms a 250~nm wide constriction. }
	\label{fig01}
\end{figure*}

\section{Sample fabrication}
All samples are fabricated using the well-established dry delamination van-der-Waals stacking technique\cite{Wan13,Bansz15}. A few-layer MoS$_2$ crystal gets partially encapsulated between two hBN flakes and is then placed onto a Si$^{++}$/SiO$_2$ substrate (Figure~1a), where the highly doped Si$^{++}$ substrate is used as a global back gate. In order to electrically contact the MoS$_2$ from the top surface, the left and the right parts of the MoS$_2$ flake remain uncovered by the upper hBN crystal. At the same time, the inner fully encapsulated area of MoS$_2$ is well protected against transfer related degradation and contamination ensuring a high quality of the device. Thereafter, one large flake of few-layer graphene (FLG) is placed on top of the entire structure, which is in direct contact with MoS$_2$ in all parts that are not protected by the upper hBN flake (Figure~1b). Figure~1d shows an optical microscopy image of such a complete van-der-Waals stack. In contrast to previous reports\cite{Roy14,Yu14,Cui15}, where multiple flakes of FLG have been transferred to fabricate electrical contacts, we only transfer a single large FLG flake that is subsequently patterned (Figure~1c). Using a single flake to contact multi-terminal MoS$_2$ devices is convenient, as it saves several transfer steps or crystal phase engineering of the MoS$_2$\cite{Kap14}. Furthermore, it prevents any direct contact of solvents or wet chemistry with the MoS$_2$. Only after the last transfer step, we immerse the sample in solvents, where also the outer parts of the MoS$_2$ are already fully encapsulated by the FLG. Electron-beam lithography and reactive ion etching (RIE) using an Ar/O$_2$ plasma with a flux of 32 sccm Ar and 8 sccm O$_2$ and a power of 60 W allow to selectively etch the FLG and the MoS$_2$, while the etching rate is negligible for the top hBN. This enables us (i) to remove the FLG covering the top hBN and (ii) to etch trenches through the FLG/MoS$_2$ heterostructure. The trenches are needed to separate the individual leads on each side of the multi-terminal devices (see illustration in Figure~1g). In a last step, Ti/Au contacts (typical thickness of 5/95 nm) to the FLG, as well as the shadow mask structures for the constriction on the top hBN (see also Figure 1c) are evaporated, followed by lift-off.

\begin{figure*}[t!]\includegraphics[draft=false,keepaspectratio=true,clip,width=0.94\linewidth]{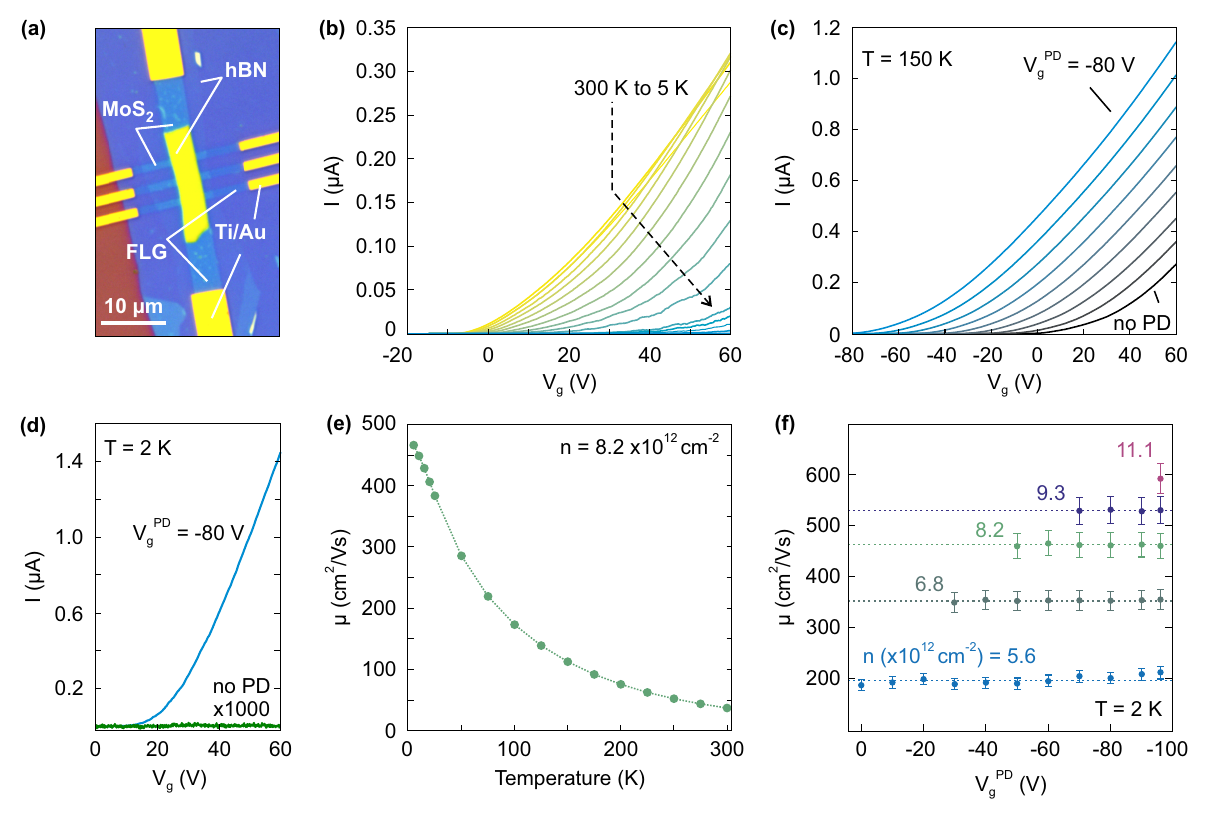}
	\caption[fig02]{(a) Optical micrograph of a Hall bar device fabricated from a hBN/trilayer MoS$_2$/hBN/FLG heterostructure. (b) Current through a Hall bar device as function of back gate voltage for temperatures ranging from 300~K to 5~K in steps of 25~K between 300~K and 25~K and steps of 5~K between 25~K and 5~K. (c) Current as function of the gate voltage in a Hall bar for several consecutive photodoping (PD) steps of $\Delta$V$^{\mathrm{PD}}_\mathrm{g}=-10$~V. (d) Current as function of gate voltage at T=2~K before and after the photodoping at  V$^{\mathrm{PD}}_\mathrm{g}=-80$~V. (e) Charge carrier mobility of a five-layer MoS$_2$ Hall bar as function of temperature, showing increased carrier mobility at low temperatures after photodoping at 150~K. (f) Charge carrier mobility as function of the photodoping gate voltage for different charge carrier densities, indicating that photodoping does not influence the charge carrier mobility of the device. } \label{fig02}
\end{figure*}

The devices are characterized by scanning confocal Raman microscopy\cite{Graf07} which enables us to spatially probe the structural (and partly electronic) quality of all individual layers of the van-der-Waals heterostructure. Additionally, we follow ref. \citenum{XZhang2015} to determine the number of layers of the MoS$_2$ from these Raman spectra. Figure~1e depicts two typical Raman spectra for a trilayer MoS$_2$ constriction device. The upper spectrum in Figure~1e is recorded in the lead area (see open circle in Figure~1f), where the MoS$_2$ is covered with FLG. The G peak at 1583~cm$^{-1}$ (purple) originates from the FLG\cite{Dressel09}. The hBN peak is located at around 1365~cm$^{-1}$ (green)\cite{Gor11} while the A$_{1g}$ and the E$^1_{2g}$ modes of the trilayer MoS$_2$ are located around 409~cm$^{-1}$ (blue) and 386~cm$^{-1}$, respectively\cite{Sah13,Ton13}. The lower spectrum in Figure~1e is recorded on the hBN/MoS$_2$/hBN area of the device, close to the shadow masks, where only the two peaks from the hBN and the MoS$_2$ are visible (see open square in Figure~1f). Figure~1f shows a Raman map of the intensities of the characteristic Raman peaks of the 2D materials in use. Green represents the intensity of the hBN peak, purple depicts the G-peak intensity of the FLG, blue shows the intensity of the A$_{1g}$ mode of the MoS$_2$. The dark areas are the gold shadow masks, which define the constriction in the underlying MoS$_2$ layer. Figure~1h shows a false color scanning electron microscopy image (SEM) of a typical device. The boundaries of the individual flakes are indicated by the dashed colored lines, where we use the same color code as in Figure~1e. Figure~1i shows a close up SEM image of a typical shadow mask structure with a width of 250~nm and a length of 175~nm.

\section{Results and discussion}
We start by discussing charge transport in a trilayer MoS$_2$ Hall bar structure and explore the method of photodoping before discussing transport through the MoS$_2$ constrictions. The Hall bar device (see Figure~2a) was fabricated the same way as described above. In Figure~2b we show the current $I$ flowing through the Hall bar, which is measured  in standard four-terminal geometry in longitudinal configuration as function of the applied gate voltage $V_\mathrm{g}$ for temperatures ranging between 300~K to 5~K at a constant bias voltage of V$_\mathrm{b}$~=~100~mV. Prior to photodoping (PD), the overall current strongly decreases with decreasing temperature and falls below our detection limit for temperatures less than 10~K (see corresponding green curve taken at 2~K in Figure~2d). The strong decrease of current at low temperatures is due to (i) a strong increase of the MoS$_2$ resistance and partially (ii) to an increase of the contact resistance (see supplementary materials, Figures S1a and S1b). The increase in resistivity of the MoS$_2$ is most likely due to localization effects. At higher temperatures, we observe a gate dependent insulator-to-metal transition identified by the onset of the current flow in Figure~2b, which continuously shifts to smaller gate voltages with increasing temperature. Similar observations of a charge carrier density dependent insulator-to-metal transition in MoS$_2$ have been recently reported\cite{Rad13,Bau13,Sch14}. Metallic conductance at low temperatures can be observed for high charge carrier densities around $n\approx10^{13}$~cm$^{-2}$. In order to adjust the charge carrier density in our devices we make use of the recently reported photodoping mechanism across a hBN-to-graphene interface\cite{Ju14,Wong15,Neu16}. Nitrogen vacancies and/or carbon impurities in the hBN crystal can be optically activated and act as a charge reservoir.
By illuminating our hBN/MoS$_2$/hBN Hall bar using a high energy light emitting diode (LED) with a center wavelength of 470~nm, these hBN defect states get optically excited. By applying a gate voltage, the activated charges are transferred into the MoS$_2$ layer leaving behind oppositely charged states in the hBN layer. The charge transfer continues until the applied (back gate induced) electric field gets fully screened by the charged hBN layer. When turning off the LED, the charges become trapped in the hBN resulting in a constant carrier doping of the illuminated area of MoS$_2$. In agreement with earlier work on similar photodoping of graphene\cite{Ju14,Neu16}, we observe that the doping is stable over the entire measurement period (up to several weeks) in the temperature range between 2~K and 250~K. All transport measurements have been performed after the LED has been switched off. As any excitation of charge carriers due to the illumination and their relaxation happens in MoS$_2$ on the order of ps to ns \cite{Shi2013,Palummo2015} we consider the effect of the illumination only to be relevant for the photodoping process.

    \begin{figure}[b!]\centering
	\includegraphics[draft=false,keepaspectratio=true,clip]{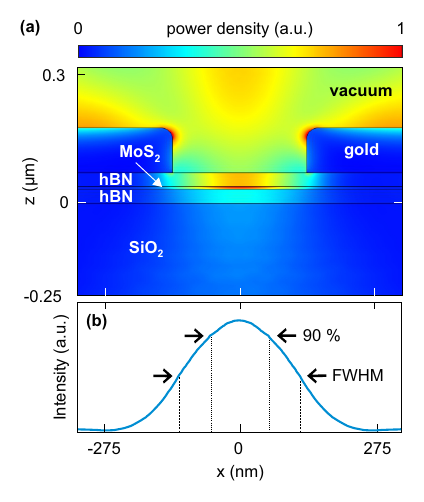}
	\caption[fig03]{(a) Simulation of the intensity distribution in the constriction region defined by a shadow mask assuming vertical illumination of the sample. (b) Intensity profile of a cross-section through the bottom hBN for 275~nm spaced shadow masks. The full-width-at-half-maximum yields a width of 240~nm. If we assume that we need 90\% of the light intensity for forming the constriction the width of it would reduce to around 120~nm or below.}
	\label{fig03}
\end{figure}

In a control experiment, where a five-layer MoS$_2$ flake is resting directly on SiO$_2$, (see supplementary materials Figure~S3) we observe a volatile and non-systematic photodoping effect, which is either related to the charge traps in the silicon oxide layer or at the SiO$_2$/graphene interface\cite{DuckKim13,Yurg13}. In contrast to the photodoping on hBN, the net doping shift is strongly dependent on the time of illumination and not so much on the applied back gate voltage. Increasing the voltage mainly leads to an increased efficiency, i.e. increased speed, of the doping process consistent with previous studies for photodoping of graphene on SiO$_2$\cite{DuckKim13}. Importantly, the induced doping is volatile, i.e. not stable over time and decays directly after switching off the LED.

\begin{figure}[t!]\centering
	\includegraphics[draft=false,keepaspectratio=true,clip,%
	width=\linewidth]{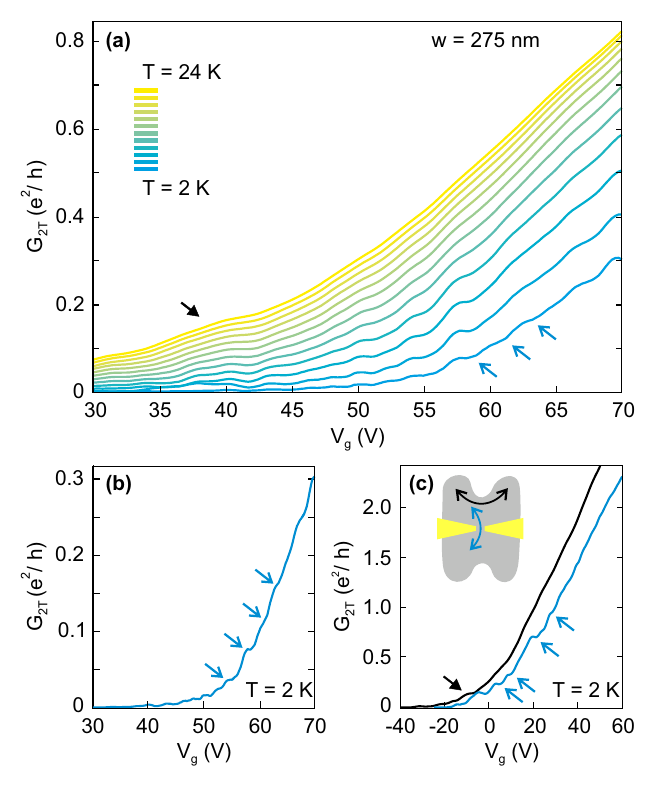}
	\caption[fig04]{(a) Two-terminal conductance G$_{2T}$ as function of gate voltage for different temperatures in steps of 2~K for a bilayer MoS$_2$ constriction. Step-like features in the conductance are observed at low temperatures.  These features vanish at 15~K, as the MoS$_2$ underneath the metallic shadow mask becomes conductive. (b) Two-terminal conductance through the constriction at 2~K, showing kinks in the conductance (see blue arrows). (c) Conductance alongside (black trace) and through a four-terminal trilayer MoS$_2$ constriction (blue trace) as function of gate voltage after photodoping. The conductance trace alongside the constriction are smooth, whereas only the trace through the constriction exhibits step-like features in the current. Inset: schematic of the current flow corresponding to the different conductance traces.}
	\label{fig04}
\end{figure}

 Figure~2c shows a series of gate characteristics of the trilayer Hall bar at 150~K before and after photodoping. The device was illuminated (photodoped) for 10 minutes at constant gate voltages V$^{\mathrm{PD}}_\mathrm{g}$ varying between V$^{\mathrm{PD}}_\mathrm{g}=-10$~V and V$^{\mathrm{PD}}_\mathrm{g}=-80$~V in steps of $\Delta$V$_\mathrm{g}=-10$~V. The illumination time is chosen in such a way that the the photodoping process is saturated. The gate characteristics show a respective overall increase in current, which corresponds well to shifts of $\Delta$V$^{\mathrm{PD}}_\mathrm{g}=-10$~V in gate voltage. At the same time the gate characteristics remain identical in slope and shape indicating a constant charge carrier mobility and a homogeneous and efficient doping process throughout the entire device. After illuminating the Hall bar at V$^{\mathrm{PD}}_\mathrm{g}=-80$~V, the device shows metallic behavior, i.e. it remains well conductive at 2~K when applying moderate gate voltages (see Figure~2d and supplementary materials Figure~S2). Charge carrier mobilities extracted from Hall measurements on a typical five layer Mo$S_2$ Hall bar device show values on the order of $\mu$~=~600~cm$^2$/(Vs) at low temperatures (2~K), which decreases to a value of around 45~cm$^2$/(Vs) at room temperature most likely due to electron-phonon scattering\cite{Rad13} (see Figure~2e). In the Figure~2f, the extracted charge carrier mobility at 2~K is plotted for different photodoping steps. For each photodoping step the device is warmed up to 150~K (as the LED is freezing out at low temperatures), the photodoping is performed and the device is cooled down again to 2~K to extract the low temperature carrier mobility. The different colors represent different charge carrier densities. The photodoping significantly increase the available range of charge carrier density $n$ as in the non-photodoped device $n$ was limited to 5.6$\times$10$^{12}$ cm$^{-2}$ when applying reasonable gate voltages. Similar to remote doping in 2DEGs\cite{Ding1978}, the mobility of the device is not affected by the process of photodoping, as the charge that is introduced in the hBN flake is spatially separated from the MoS$_2$ (see colored dashed lines in Figure~2f). Additionally, we do not observe any significant change of $n$ at a fixed gate voltage with respect to the temperature after the photodpoing up to T=150 K. In order to underline the high electronic quality of the heterostructures, we also measure the quantum Hall effect in the five layer MoS$_2$ Hall bar\cite{Cui15}, where we extract quantum mobilities on the order of 2,000~cm$^2$/(Vs) (see supplementary material Figure~S3).

Next we discuss constriction devices. For observing size-quantization effects, including quantized conductance, the Fermi wavelength of the carriers has to be on the order of the constriction width. For reasonable carrier densities this requires devices with feature sizes on the order of 100~nm (or below). In order to harness photodoping for such small device feature sizes, we employ a shadow masking technique using non-transparent metal structures.
Notably, this approach is in contrast to previous work, where some of us demonstrated photodoping in graphene/hBN heterostructures with micron-scale spatial resolution using a confocal laser set-up\cite{Neu16}. The metallic shadow masks, predefined by electron beam lithography, allow to define long-lasting doping profiles in arbitrary geometries (see e.g. Figures~1g-1i). Using this shadow masking technique, we illuminate only the source and drain contact areas (leads) and the constriction, while everything covered by the metal masks is unaffected by the photodoping process. As the MoS$_2$ is non-conductive at 2~K prior to the photodoping no depletion is necessary to form a constriction between the metallic shadow masks. In contrast to conventional constrictions in 2DEGs, we neither have to etch our sample nor have to use electrostatic gates to define the structure as free carriers are only available in the regions, where the sample has been photodoped. In the following, the metallic shadow mask is only used to define the constriction by the photodoping. The devices are only tuned by applying a gate voltage V$_g$ to the highly doped Si$^{++}$ back gate.

In order to estimate the expected width of the constriction we perform numerical simulations\footnote{The simulation was performed using the COMSOL Multiphysics package and assumes vertical illumination of the sample.} of the (lateral) light intensity profile responsible for forming the constriction during photodoping. Figure~3 shows the simulated intensity distribution in the constriction region defined by a 275~nm spaced shadow mask. The simulation indicates that the light intensity underneath the shadow mask is strongly suppressed, with a lateral extend of the intensity profile at full-width-at-half-maximum (FWHM) of 240~nm (see arrows in Figure~3b). It has to be emphasized that this value can only be regarded as an upper limit as we do not know (i) the intensity at which photodoping becomes efficient, (ii) to which extend electrostatic effects from the metallic shadow mask and (iii) additional plasmonic effects between the shadow masks may affect the potential landscape in the constriction region. When assuming 90\% of the light intensity is needed to form the constriction its width would reduce to around 120 nm or below. To further investigate the influence of these effects, additional experiments are necessary which are beyond the scope of this work. Similar simulations have been performed for different constriction width showing similar results.

Figure~4a shows the two-terminal conductance through a 275~nm wide constriction fabricated from a bilayer MoS$_2$ flake at constant bias of V$_\mathrm{b}$~=~3~mV for temperatures ranging from 2~K to 24~K (after photodoping the device at 50~K and V$^{\mathrm{PD}}_\mathrm{g}=-60$~V). The conductance trace at 2~K shows step-like features suggesting confinement effects in the constriction (see Figure~4b). A very similar behavior is observed for all five measured constriction devices (see supplementary Figure~S5). The step-like features in the conductance traces remain visible up to 15~K (Figure~2a) with increasing two-terminal conductance. The increasing conductance with a fixed position of the features with respect to V$_g$ could well be a sign of increasing transmission through the constriction until the confinement gets destroyed when the MoS$_2$ underneath the metallic shadow masks starts to contribute to the transport.

\begin{figure}[tb]\centering
	\includegraphics[draft=false,keepaspectratio=true,clip,%
	width=\linewidth]{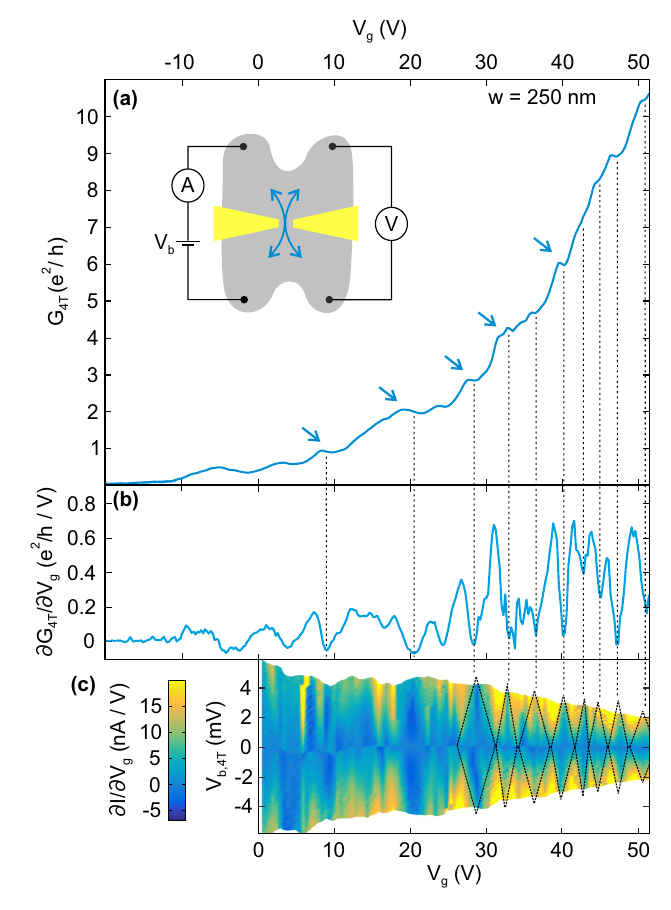}
	\caption[fig05]{(a) Four-terminal conductance as function of the applied gate voltage at 2~K for V$_\mathrm{b}=7$~mV showing step-like features which are roughly spaced by one $e^2/h$. (b) Derivative with respect to the applied gate voltage of the trace shown in (a), showing dips that correspond to the steps in (a). (c) Bias spectroscopy measurement showing diamonds at the positions of the steps, indicated by the black dashed lines.}
	\label{fig05}
\end{figure}

In order to verify that the conductance steps originate from the constriction, we perform two-terminal measurements in different contact configurations on a separate multi-terminal device fabricated from a trilayer MoS$_2$ flake (see Figure~4c) with a constriction width of 250~nm. The black trace in Figure~4c shows the conductance between two contacts on one lead, i.e. on one side of the constriction of a four-terminal device, (see the black arrows in the inset of Figure~4c). The measurements were performed at 2~K after photodoping the device at V$^\mathrm{PD}_\mathrm{g}=-60$~V and 50~K. The trace shows a smooth and monotonic increase of the conductance with the applied gate voltage except for a feature close to the onset of the conductance at around V$_\mathrm{g}=-5$~V (see black arrows) and is overall very similar in shape to the traces obtained from several Hall bar devices. The feature close to the transport onset is most likely related to localized states due to the random potential variations from trapped charges\cite{Ghat11}. In contrast, the conductance through the constriction (blue trace in Figure~4c) shows the similar extended step-like features as they were observed for the two-terminal device.

\begin{figure}[b]\centering
	\includegraphics[draft=false,keepaspectratio=true,clip,%
	width=\linewidth]{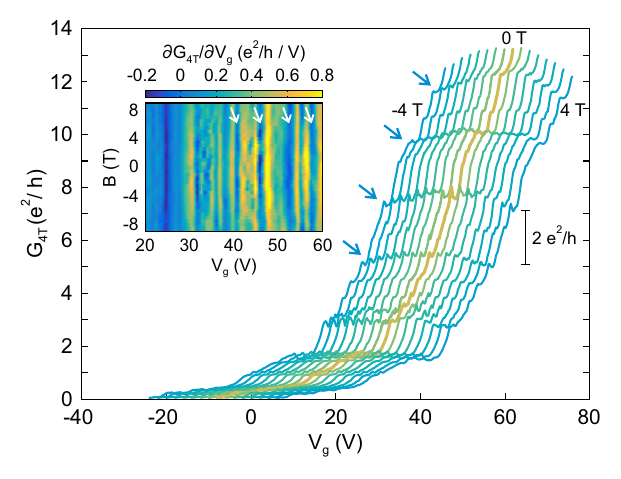}
	\caption[fig06]{ Four-terminal conductance as function of the applied gate voltage at 2~K for a constant bias of V$_\mathrm{b}$~=~-6~mV at fixed perpendicular magnetic fields B ranging from -4~T to 4~T for the same device shown in Figure~5 after a second cool-down. Again we observe step-like features which are roughly spaced by one $e^2/h$. The inset shows the transconductance dG/dV$_g$ as a function of the applied gate voltage and the magnetic field for V$_\mathrm{b}$~=~-6~mV. The white arrows highlight the same features marked by the blue arrows in the main figure.}
	\label{fig06}
\end{figure}

The total resistance from the two-terminal data shown in Figure~4 heavily depends on the particular device. It ranges between 10 and 100~k$\Omega$ and is thus significantly larger than the conductance quantum, as the contact resistances heavily contributes to the overall resistance. A subtraction of these parasitic resistances proves difficult as both are strongly gate voltage and bias dependent. In order to reduce the influence of these additional device resistances on the measurement, we performed four-terminal measurements on the multi-terminal trilayer MoS$_2$ device (see inset of Figure~5a). It is important to point out that this four-terminal measurement allows to extract an effective two-terminal conductance as the voltage is probed roughly 4~$\mu m$ away from the constriction. This length is significantly larger than the elastic mean free path in the MoS$_2$ 2DEG. 
In Figure~5a we show the four-terminal conductance $G_{4T}$ in unit of the conductance quantum $e^2/h$ as a function of the applied gate voltage measured at 2~K. The conductance ranges from 0 to 11~$e^2/h$ and exhibits step-like features which are roughly spaced by 1 to 2 $e^2/h$ (see blue arrows). This observation is in contrast to the expected appearance of steps at multiples of 6~$e^2/h$ for a narrow constriction in trilayer MoS$_2$ due to the six-fold degeneracy of the three $Q$ and three $Q^\prime$ valleys, which govern the transport\cite{ZWu16,Piso17}. However, it is in reasonable agreement with recent measurements performed on a gate-defined nanoconstriction in a MoS$_2$-based heterostructure\cite{Piso17}. It might be explained by lifting of the level degeneracies due to the additional confinement of the constriction and/or a reduced transmission through the constriction while we cannot completely exclude that the FLG/MoS$_2$ interface and the MoS$_2$ regions leading to the constriction cause an additional series resistance. The reduced transmission would also explain (i) that the observed step sizes do not meet precisely multiples of 1 or 2 $e^2/h$ and (ii) the substructure observed on the gate voltage derivative of $G_{4T}$ (Figure~5b) and the bias-dependent measurements (Figure~5c). The latter exhibit well developed diamond-like features for high gate voltages ($>30$~V) from which we extract an energy scale of around 3 to 4~meV for the involved subband spacings, where the degeneracy is potentially lifted. At lower carrier density (smaller gate voltages) the bias spectroscopy measurements show signatures of intersecting diamonds, which we attribute to localized states forming in the narrow constriction.
In Figure~6 we show the four-terminal conductance as function of gate voltage for different perpendicular magnetic fields $B$ ranging from $B=-4$ to $4$~T. These measurements have been taken on the same device shown in Figure~5 but at a different cool-down. The traces are offset by V$_\mathrm{g}=\pm2$~V for clarity. We again observe well-developed step-like features with a step height on the order of  1 or 2 $e^2/h$ (see e.g. the step from $G=5$ to 7 $e^2/h$ at 4 T). Most importantly, we observe that the step-like features are virtually not tuned by the applied $B$-field. This, on one hand, rules out universal conductance fluctuations as a possible explanation for the observed step- and kink-like features in our measurements. On the other hand, it provides strong indication that the formed constriction is rather narrow such that orbital effects are not playing an important role. The inset in Figure~6 shows the transconductance dG$_{4T}$/dV$_g$ as function of V$_g$ and $B$-field highlighting that the step-like features are indeed not shifting much with $B$-field. We do not know why the experimental data show this weak dependence and why the step-like features only qualitatively match quantized conductance. Further improvement of the device, i.e. MoS$_2$-material quality is needed for more detailed studies on quantized conductance steps in MoS$_2$ or related TMDC materials.

\section{Conclusion}
In summary, we present a technique to fabricate mesoscopic devices based on MoS$_2$ encapsulated by hBN using an entirely dry transfer process, contacting the MoS$_2$ using few-layer graphene and adjusting the charge carrier concentration using photodoping. By transferring only a single FLG flake to fabricate a large number of contacts to the MoS$_2$, we reduced the complexity of the required van-der-Waals heterostructure significantly. Similar to remote doping in 2DEGs, we are able to adjust the doping level in MoS$_2$, while preserving the charge carrier mobility, resulting in metallic transport at low temperatures. By using metal shadow masks combined with photodoping, we demonstrate a way to produce complex doping profiles on a scale close to the diffraction limit, allowing to observe quantum transport in MoS$_2$ constrictions. In low temperature transport measurements, we reproducibly observe signatures of quantum confinement effects in MoS$_2$ constrictions. Even though gaining precise control over the exact confinement potential proves still to be difficult, our simulations and measurements show that the shadow masking technique can prove useful to generate smooth and long-lasting doping profiles in van-der-Waals heterostructures although most likely the lateral size of structures induced by photodoping is limited by the diffraction limit. This technique could pave the way towards the fabrication of more complex mesoscopic devices in TMDCs, such as for example Aharonov-Bohm rings or antidot lattices, which are difficult to realize using electrostatic gating.

\section{Acknowledgement}
The authors thank F.~Haupt for fruitful discussions and 
S.~Staacks for help with the figures. Support by the Helmholtz Nano Facility (HNF)\cite{HNF17}, the DFG (SPP-1459), the ERC (GA-Nr. 280140), and the European Union's Horizon 2020 research program under grant agreement No. 696656 (Graphene Flagship) are gratefully acknowledged. Growth of hexagonal boron nitride crystals was supported by the Elemental Strategy Initiative conducted by the MEXT, Japan and JSPS KAKENHI Grant Numbers JP26248061, JP15K21722 and JP25106006.

\end{document}